\documentstyle[preprint,prl,aps]{revtex}

\begin{document}
\draft

\title{
Correlations in a two--chain Hubbard model
}
\author{ R.M.\ Noack and S.R.\ White }
\address{
Department of Physics
University of California
Irvine, CA 92717
}
\author{ D.J.\ Scalapino }
\address{
Department of Physics
University of California
Santa Barbara, CA 93106
}
\date{\today}
\maketitle
\begin{abstract}
Equal time spin--spin and pair field correlation functions are
calculated for a two-chain Hubbard model using a density--matrix
numerical renormalization group approach.
At half--filling, the antiferromagnetic
and pair field correlations both decay
exponentially with the pair field having a much shorter correlation
length.
This is consistent with a gapped spin-liquid ground
state.
Below half--filling,  the antiferromagnetic correlations
become incommensurate and the spin gap persists.
The pair field
correlations appear to follow a power law decay which is similar to
their non-interacting $U=0$ behavior.
\end{abstract}


\narrowtext

Materials such as $(\text{VO})_2 \text{P}_2 \text{O}_7$ \cite{johnston} and
$\text{Sr}_2 \text{Cu}_4 \text{O}_6$ \cite{takano,rice1} contain weakly
coupled arrays of metal-oxide-metal ladders.
In the nominal insulating state, the metal ions have spin one-half and
the oxygens mediate an antiferromagnetic superexchange coupling.
Calculations for a two-chain antiferromagnetic model with an exchange
interaction $J$ along each chain and $J'$ across each rung have shown
that the insulating state can have a spin gap
\cite{dagotto,strong,barnes,parola}.
A $t$--$J$, $t'$--$J'$ model was introduced to describe the hopping of
holes in the doped system.
Lanczos calculations for this model with
$J' > t'$ found enhanced pairing
correlations on a $2\times 8$ ladder doped near one-quarter
filling \cite{dagotto}.
A mean--field analysis of this model using a Gutzwiller
renormalization of the matrix elements \cite{rice2} found a spin gap
in the insulating state which initially increases with doping.
In addition, the doped system near half-filling was found to have a
mean--field superconducting order parameter with a modified d--wave
structure.

Another representation of coupled chains is provided by the two-chain
Hubbard model \cite{noack}.
Here we report results obtained for the ground-state magnetic and pairing
correlations of the two--chain Hubbard model using a
density--matrix numerical renormalization group approach \cite{white}
to study lattices of up to $2\times 32$ sites.
We find that at half--filling, the dominant correlations are
antiferromagnetic, but that these decay exponentially because of a
spin gap, which we calculate directly.
The singlet pair field correlations also decay exponentially but with
a short correlation length of the order of the lattice spacing.
The ground state at half-filling is thus a spin liquid.
As the system is doped away from half-filling,
the antiferromagnetic correlations show an incommensurate structure
and a peak develops in the magnetic structure factor
at a wave vector proportional to the filling.
The spin gap decreases as the system is doped, but persists down to an
occupation of 0.75 electrons per site or less.
The singlet pair field correlations are enhanced
and exhibit a power--law decay, but the form of the decay is approximately
$\ell^{-2}$, where $\ell$ is the separation distance.
This is the
same power-law dependence as for the noninteracting, $U=0$, system.

We picture the two--chain Hubbard model as a ladder standing along the
$y$--axis with its rungs along the $x$--axis so that
\widetext
\begin{equation}
H=
- t_y \sum_{ i, \lambda \sigma}
( c^{\dagger}_{i,\lambda \sigma}c_{i+1,\lambda \sigma} +
 c^{\dagger}_{i+1,\lambda\sigma}c_{i,\lambda \sigma}  )
- t_x \sum_{i, \sigma}
( c^{\dagger}_{i,1 \sigma}c_{i,2 \sigma } +
 c^{\dagger}_{i, 2\sigma}c_{i, 1 \sigma} )
+ U\sum_{ i, \lambda}
n_{i,\lambda \uparrow}n_{i,\lambda \downarrow }.
\end{equation}
\narrowtext
Here $c^{\dagger}_{i, \lambda \sigma}$ creates an electron of spin
$\sigma$ at rung $j$ and side
$\lambda=1$ (left) or $2$ (right),
the hopping along a chain is $t_y$, the hopping between chains on a
rung is $t_x$, and $U$ is the on--site Coulomb repulsion.

The density matrix formulation of the renormalization group
\cite{white} provides a controlled approximation for the calculation
of ground--state energies and correlation functions.
We have used the finite--system method with open boundaries
and have studied $2 \times 16$, $2 \times 24$, and
$2 \times 32$ lattices, keeping up to 400 states per block.
Details of the calculations will be published elsewhere.
The discarded density matrix weight (truncation error) varies from
$5.5 \times 10^{-5}$ to less than
$10^{-9}$ for the $2\times 32$ results shown here.
In addition, we have examined the convergence of measured quantities
as a function of the number of states kept and ascertained that the
symmetries of the Hamiltonian are preserved.
The maximum errors in the quantities shown here are at
most a few percent, and in most cases are much smaller.
Since the method works best with chains that have open boundaries at the
ends, all results here have open boundaries.

We have calculated the equal-time spin and pair field correlation
functions
$S_{\lambda\lambda '}(i,j)=\langle M^z_{i,\lambda} M^z_{j,\lambda '}\rangle$,
$D_{xx}(i,j)= \langle \Delta_{x i} \Delta^\dagger_{x j} \rangle$,
and
$D_{yx}(i,j) = \langle \Delta_{y i} \Delta^\dagger_{x j} \rangle$
with
\begin{eqnarray}
\begin{array}{cl}
M^z_{i,\lambda} & = n_{i,\lambda \uparrow} - n_{i,\lambda \downarrow}\\
\Delta^\dagger_{x i} & = c^\dagger_{i,1 \uparrow} c^\dagger_{i,2 \downarrow}
- c^\dagger_{i,1 \downarrow} c^\dagger_{i,2 \uparrow}  \\
\Delta^\dagger_{y i} & =
 c^\dagger_{i+1,2 \uparrow} c^\dagger_{i,2 \downarrow}
- c^\dagger_{i+1,2 \downarrow} c^\dagger_{i,2 \uparrow}. \\
\end{array}
\end{eqnarray}
Here $S_{11}(i,j)$ and $S_{12}(i,j)$ measure the spin-spin
correlations along a chain and between the chains respectively,  and
$D_{xx}(i,j)$ measures the singlet pair field correlations in which a
singlet pair is added at rung $j$ and removed at rung $i$.
In addition, $D_{yx} (i,j)$ measures the pair field correlations in
which a singlet pair is added to rung $j$ and removed from the
right--hand chain between rungs $i$ and $i+1$.
The relative phase of the pair wave function across the $i$th rung to
along one chain from $i$ to $i+1$ is given by comparing the phase of
$D_{xx}(i,j)$ to $D_{yx}(i,j)$.
This turns out to be negative, corresponding to the mean field result
obtained in Ref. \cite{rice2}.
However, the non-interacting $U=0$ result at a filling
$\langle n \rangle = 0.875$ is also negative.

Because of the open boundaries at the ends,
the system is not translationally invariant.
Therefore, the correlation
functions are dependent on both $i$ and $j$, whereas in the
thermodynamic limit or with periodic boundary conditions, they are a
function of only $| i-j |$.
We have found that the spin correlations at all fillings and the
pairing correlations at half--filling are not strongly dependent on
the placement of $i$ and $j$, so we choose them to be as symmetric
about the center of the lattice as possible, in order to minimize end
effects.
The pair correlations away from half--filling have strong variation
with lattice placement, so in order to minimize these effects, we
average over a number of $i$ and $j$ for a given $|i-j|$.
We will then discuss the
correlation functions as functions of $\ell \equiv | i-j |$ with these
procedures implicit.
As we shall see, there will be discernible boundary effects, but the
lattices are long enough that one can still extract the general
behavior.
We also discuss the average filling, defined as
$\langle n \rangle = 1/N\sum_{i,\lambda \sigma} n_{i,\lambda \sigma}$
where $N$ is the number of lattice sites.

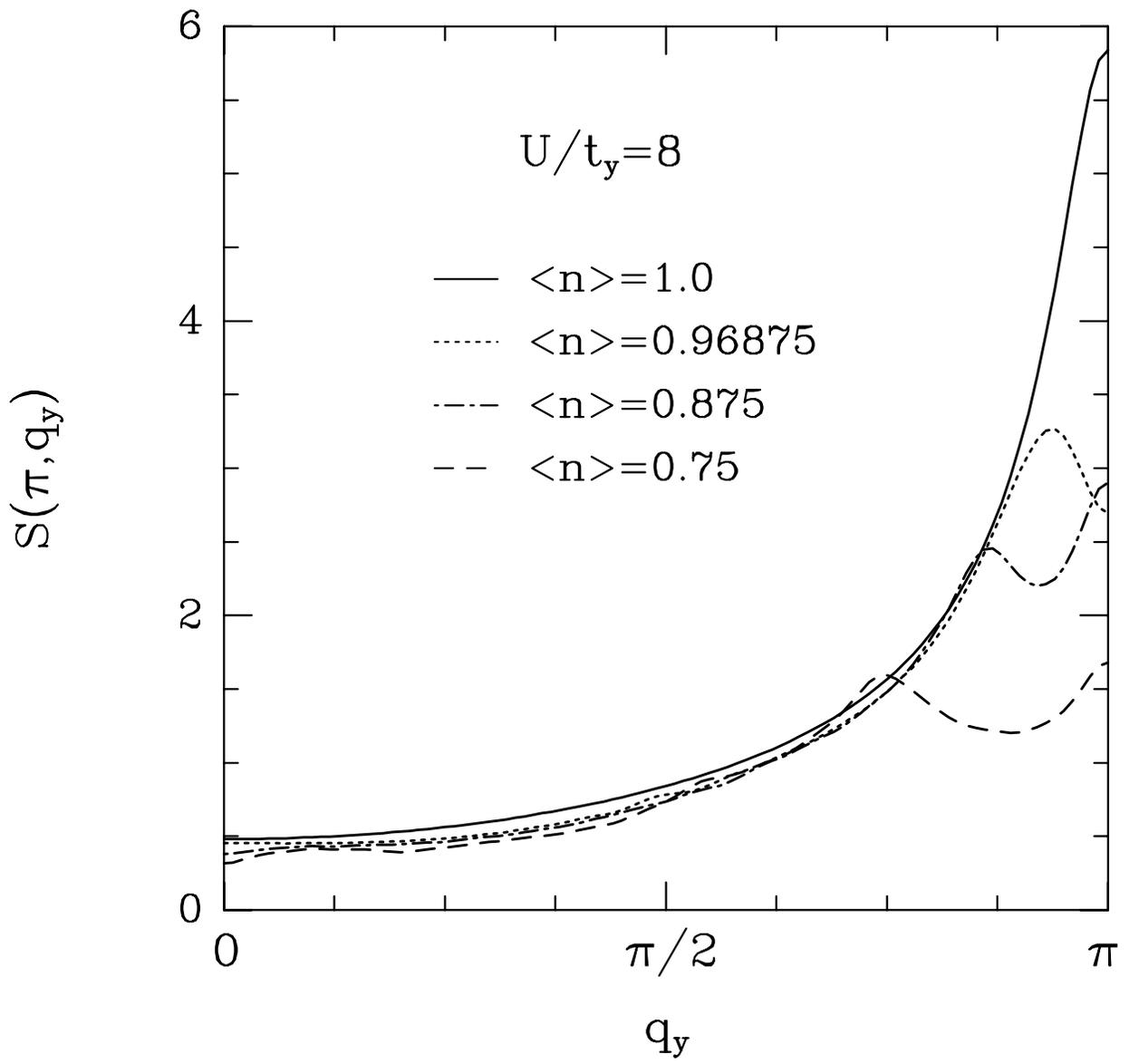
\begin{figure}
\caption{
The fourier transform $S(\pi,q_y)$
of the spin--spin correlation function $S_{\lambda \lambda'}(\ell)$.
Here $t_x=t_y$ and $U=8t_y$ and the calculations were made on a
$2 \times 32$ lattice.
}
\label{figSq}
\end{figure}

In order to understand the nature of the spin--spin correlations
we have calculated the magnetic structure factor $S(q_x,q_y)$ by
taking the fourier transform of $S_{\lambda\lambda '}(i,j)$.
Since there are two chains, $q_x$ can take on only the values $0$
and $\pi$.
For the purposes of the fourier transform, we take
$S_{\lambda\lambda'}(\ell) = 0$ for $\ell$ larger than the lattice
size.
This introduces only a small error since the
$S_{\lambda\lambda'}(\ell)$ decays exponentially with $\ell$, and has
decayed by a factor of at least 50 at the maximum lattice separation.
The correlation function $S(\pi,q_y)$ is shown for four different
fillings, $\langle n \rangle = 1$, $\langle n \rangle = 0.9875$,
$\langle n \rangle = 0.875$, and
$\langle n \rangle = 0.75$ in Fig. \ref{figSq}.
These fillings correspond to doping the half--filled system with 0, 2,
8, and 16 holes.
At $\langle n \rangle = 1$, there is a single strong peak at $q_y=\pi$.
As the system is doped away from half--filling, the peak at $q_y=\pi$
is strongly suppressed and $S(\pi,q_y)$ peaks at
$q_y = \langle n \rangle \pi$.
The residual peak at $q_y = \pi$ is present only for even numbers of
hole pairs.
Therefore, we believe that it is a finite size effect and will
vanish in the thermodynamic limit.
Since the correlations are strongly antiferromagnetic across the
chain, $S(0,q_y)$ (not shown) is small, flat and does not change much
with the filling.

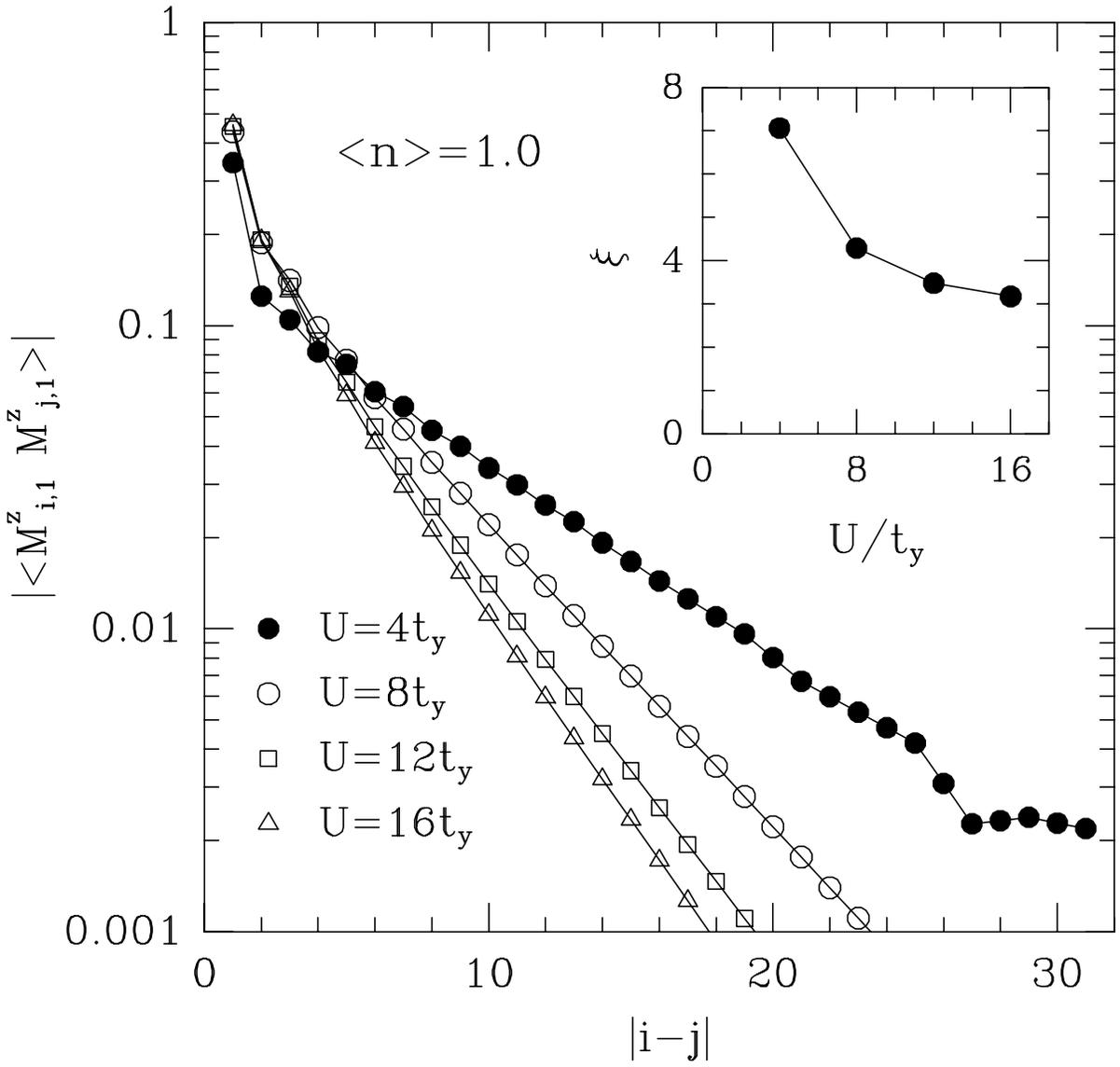
\begin{figure}
\caption{
A semilog plot of the spin--spin
correlation function for $\langle n \rangle = 1$ as a
function of $U/t_y$  for $t_x=t_y$ on a $2 \times 32$ lattice showing
the exponential decay of these correlation functions.
The inset shows the correlation length in units of the lattice
spacing taken from the slopes of the lines.
}
\label{figlS11}
\end{figure}

At half--filling, $\langle n \rangle =1$, the magnetic structure
factor shown in Fig.\ \ref{figSq} has a Lorentzian line shape
corresponding to an exponential decay of the spin--spin correlations.
This is illustrated in Fig.\ \ref{figlS11}, in which the logarithm of
$S_{11}(\ell)$ is plotted versus $\ell$ for various values of $U$.
The oscillatory deviations from linear behavior at large separation
are due to the effects of the open end boundary conditions.
{}From the slopes of the straight line segments, we have determined the
correlation length $\xi$ and this is plotted versus $U/t_y$ in the
inset of Fig.\ \ref{figlS11}.
As $U/t_y$ increases, $\xi$ decreases, saturating at a value of order 3
lattice spacings for $t_x = t_y$.
The interchain spin--spin correlations $S_{12}(\ell)$ decay
with the same correlation length as $S_{11}(\ell)$.
At half--filling, the pair field correlations also decay
exponentially, but with a correlation length
of order the lattice spacing or smaller.
Thus the pair field correlations are negligible at half--filling.

We have also carried out calculations for $U=8t_y$ at a cross--chain
hopping $t_x=0.5t_y$ and $1.2t_y$.
The correlation length decreases with increasing $t_x$,
ranging from 10.4 lattice spacings at $t_x=0.5t_y$, to 4.3 lattice
spacings at $t_x=t_y$ and 2.9 lattice
spacings at $t_x=1.2t_y$.
We expect the correlation length to diverge as $t_x \rightarrow 0$
because on a single chain there is no spin gap and the spin--spin
correlation function decays as a power law \cite{schulz}.
The largest spin gap and thus the smallest correlation length should
occur in the large $t_x/t_y$ limit, in which
the system can be described as decoupled spin singlets on the rungs of
the ladder \cite{dagotto}.

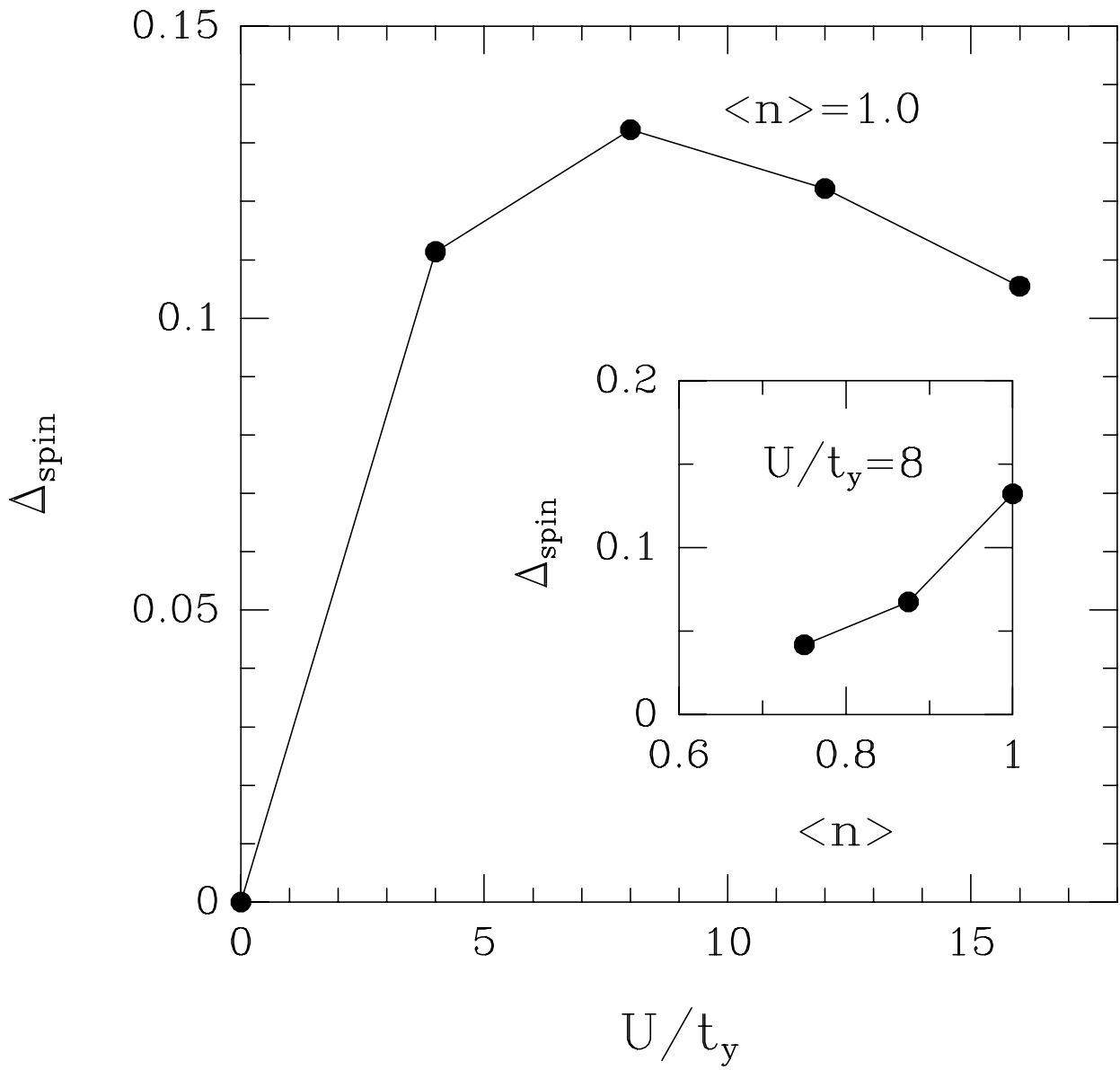
\begin{figure}
\caption{
The spin gap $\Delta_{\rm spin}$ as a function of $U/t_y$ for $t_x=t_y$
calculated on a $2\times 32$ lattice at $\langle n \rangle=1.0$.
The inset shows $\Delta_{\rm spin}$ plotted as a function of filling
$\langle n \rangle$ for $U/t_y=8$.
}
\label{figgap}
\end{figure}

In order to calculate the spin gap directly, one can calculate the
energy difference between the lowest $S_z=0$ state and the lowest $S_z=1$
state.
For the $2\times 32$ system, the energy gap at half--filling is given by
$\Delta_{\rm spin} = E(32,32) - E(33,31)$ where
$E(N_\uparrow,N_\downarrow)$ is the ground
state energy of the system with $N_\uparrow$ up electrons and
$N_\downarrow$ down electrons.
The spin gap as a function of $U$ at half--filling for $t_x = t_y$
is shown in the main plot in Fig.\ \ref{figgap} for a $2 \times 32$
lattice.
We have examined the spin gap as a function of lattice size in order
to confirm that the gap is present in the thermodynamic limit.
For $U=8t_y$, $\Delta_{\rm spin} = 0.151t_y$ on a $2\times 16$ lattice
and $\Delta_{\rm spin} = 0.132t_y$ on a $2 \times 32$ lattice.
At $U=0$, the spin gap vanishes because there are a set of
degenerate, half-occupied states on the Fermi surface.
The gap peaks at approximately $U=8t_y$ then decreases with increasing
$U$.
The gap decreases for large $U$ because it varies as $J=4t_y^2/U$
in the large $U$ limit.
In this same limit, the correlation length saturates since it varies
as the spin wave velocity divided by the gap.
The inset plot shows the spin gap as a function of filling for $U=8t_y$.
The gap decreases fairly rapidly with filling, but is still present at
$\langle n \rangle=0.75$.

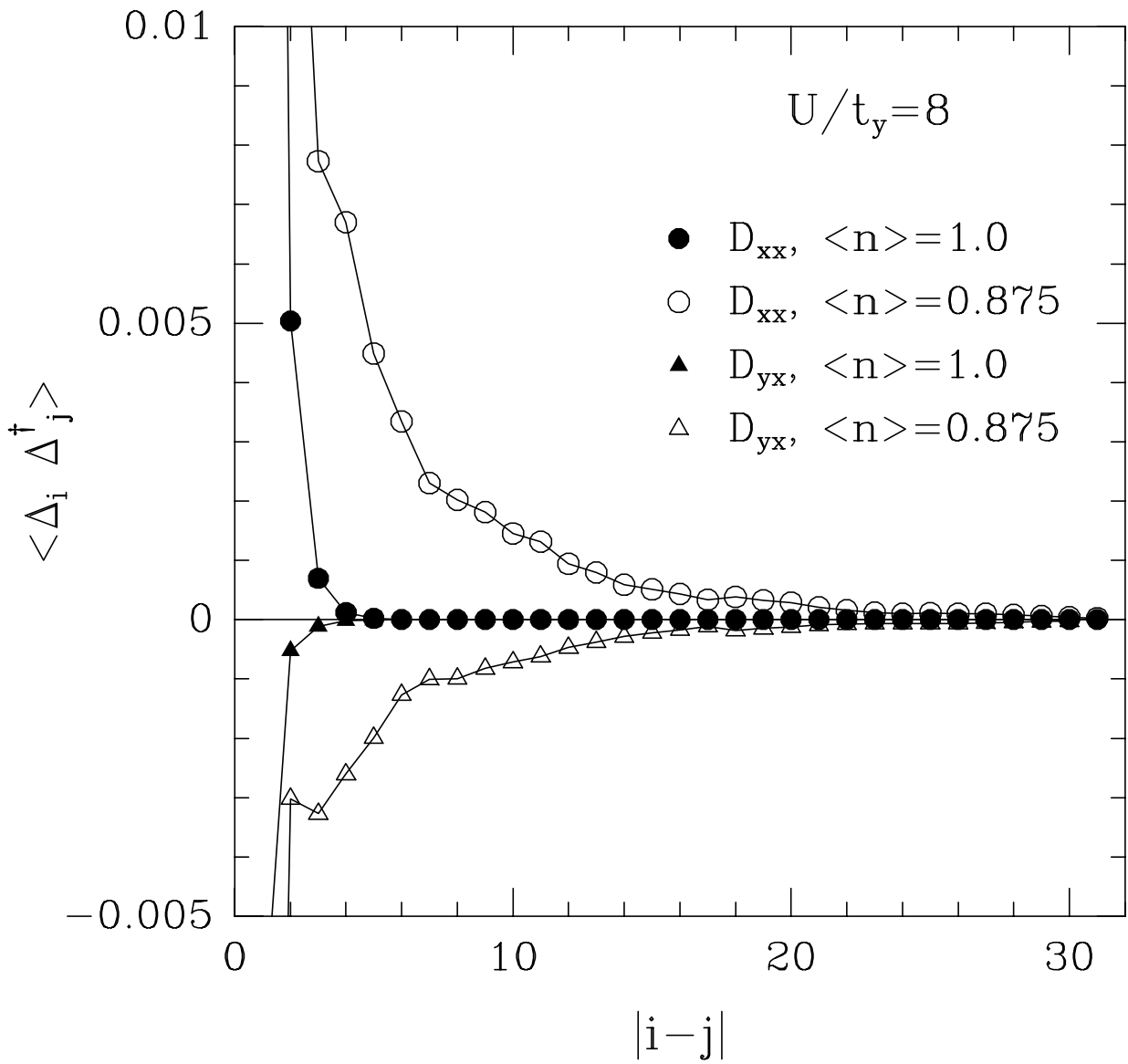
\begin{figure}
\caption{
The pair field correlation functions
$D_{xx}(\ell) = \langle \Delta_{x i} \Delta^\dagger_{x j} \rangle$ and
$D_{yx}(\ell) = \langle \Delta_{y i} \Delta^\dagger_{x j} \rangle$
versus $\ell \equiv | i-j |$ for $\langle n \rangle = 1.0$ and
$\langle n \rangle = 0.875$.
Here $t_x=t_y$ and $U=8t_y$ on a $2 \times 32$ lattice.
}
\label{figDxxyx}
\end{figure}

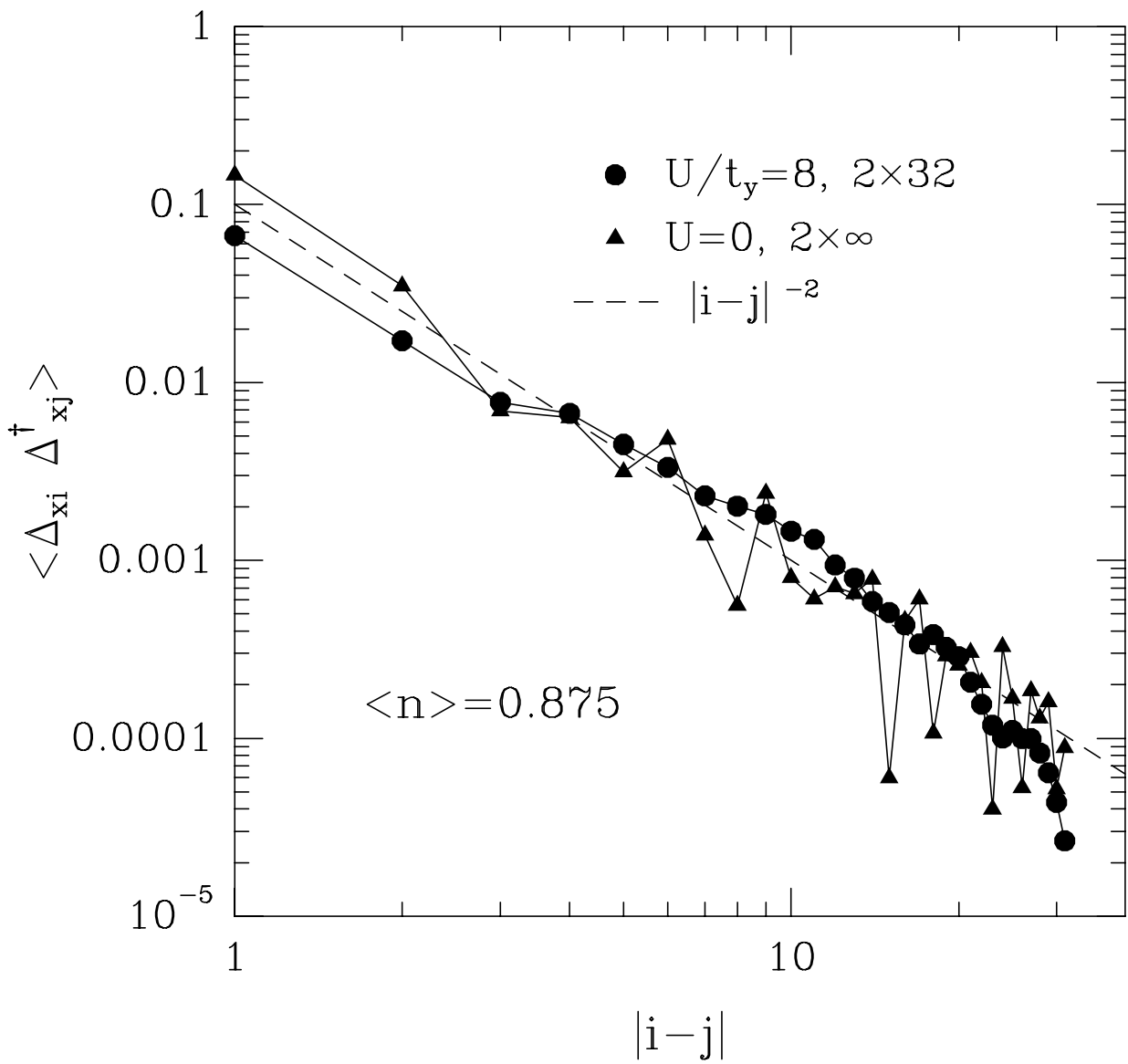
\begin{figure}
\caption{
Log-log plot of
$D_{xx}(\ell) = \langle \Delta_{x i} \Delta^\dagger_{x j} \rangle$
%and
%$\overline{D}_{xx}(\ell)$ as defined in Eq.\ (4)
versus
$l \equiv | i-j |$ at $\langle n \rangle = 0.875$.
The $U=0$ results are taken from Eq. (3)
and the dashed line has slope $-2$.
}
\label{figllDx}
\end{figure}

When the Hubbard ladder is doped away from half--filling,
($\langle n \rangle = 0.875$) one sees from Fig. \ref{figSq}
that the antiferromagnetic correlations develop an
incommensurate peak.
The pairing correlation functions $D_{xx}(\ell)$ and $D_{yx} (\ell)$ are
shown in Fig.\ \ref{figDxxyx} for $t_x=t_y$ and $U=8t_y$ at
fillings of $\langle n \rangle = 1$ and $\langle n \rangle = 0.875$.
At $\langle n \rangle = 1$, the pairing correlations decay quite
strongly compared to those at $\langle n \rangle = 0.875$.
The relative sign of $D_{xx}(\ell)$ and $D_{yx} (\ell)$ is negative at
both fillings, consistent with a modified $d$-wave structure.
In order to determine the strength of the pairing correlations,
one must consider their $\ell$--dependence at large distances.
For a one--dimensional system, we expect that any pairing correlation
will at best decay as a power of $\ell$ and can in some cases
decay exponentially, as we have seen for the half-filled system.
For two chains, one can compare with the
the non-interacting $U=0$ ladder, for which
\begin{equation}
D_{xx}(\ell) = ( 1/2 \pi \ell )^2
\left[ 2 - \cos ( 2 k_f(0)\ell) - \cos (2 k_f(\pi) \ell) \right].
\label{uzpair}
\end{equation}
Here $k_f(0) = \cos^{-1} (t_x+\mu)/2$ and $k_f(\pi) = \cos^{-1} (t_x-\mu)/2$
are the Fermi wave vectors corresponding to the bonding and
antibonding bands of the two coupled chains with $\mu$ the chemical
potential.
In order to examine the decay of the pair correlations, we have made
the log-log plot of $D_{xx}(\ell)$ shown in Fig.\ \ref{figllDx}.
We compare the $U = 8 t_y$ results with the infinite system $U=0$
results given by Eq. (\ref{uzpair}).
Our results for the interacting $2 \times 16$ and  $2 \times 24$
lattices are consistent
with Fig.\ \ref{figllDx} and thus while end effects are present, it
appears that
the equal time pair field correlations of the interacting system
decay approximately as $\ell^{-2}$.

%In order to determine the size of the pairing interaction
%vertex \cite{whitepbar},
%one can compare the full correlation function
%$D_{xx}(\ell)$ with
%\begin{equation}
%\overline{D}_{xx}(\ell) = 2 ( G_{11}^2(\ell) + G^2_{21}(\ell))
%\end{equation}
%which is constructed from the dressed single--particle Green's function
%\begin{equation}
%G_{\lambda\lambda'}(i-j) =
%-\langle c_{i,\lambda \sigma} c^\dagger_{j,\lambda' \sigma} \rangle.
%\end{equation}
%The function $\overline{D}_{xx}(\ell)$ is also plotted in
%Fig.\ \ref{figllDx} and comparing it with
%$D_{xx}(\ell)$, one sees that the effect of the interaction vertex is
%to enhance the pairing correlations.

In summary, we have found that
the Hubbard model on two chains exhibits spin--liquid behavior at
half--filling.
Both spin--spin and pairing correlations decay
exponentially with the pairing correlations having a much shorter
correlation length.
There is a spin gap which increases as the interaction $U$ is turned on,
peaks, and then becomes smaller.
When the system is doped away from half--filling, the spin gap
decreases with filling but persists down to at least
$\langle n \rangle = 0.75$.
The spin--spin
correlations become incommensurate and the $d$--wave--like pair field
correlation are enhanced.
The pair field correlations appear to decay as a power law with
exponent close to $-2$, similarly to those in the non-interacting
$U=0$ system.

We have recently received a preprint by Tsunetsugu {\sl et al.}
\cite{tsunetsugu} in which the spin gap and binding energy for two
holes is
calculated for the $t$--$J$ model on two chains via Lanczos
diagonalization.

%\section*{Acknowledgements}

The authors thank N. Bulut, T.M. Rice, A. Sandvik, M. Vekic, E.
Grannan, and R.T. Scalettar for useful discussions.
R.M.N. and S.R.W. acknowledge support from the Office of Naval
Research under grant No. N00014-91-J-1143 and D.J.S. acknowledges support
from the National Science Foundation under grant DMR92--25027.
The numerical calculations reported in this paper were performed at
the San Diego Supercomputer Center.
The computer code was written in C++.

\end{document}